\documentclass[12pt,amsfonts]{article}
\begin{document}
\def\p {{\partial}}
\def\n {{\nu}}
\def\m {{\mu}}
\def\a {{\alpha}}
\def\bt {{\beta}}
\def\f {{\phi}}
\def\th {{\theta}}
\def\g {{\gamma}}
\def\eps {{\epsilon}}
\def\e {{\psi}}
\def\la {{\lambda}}
\def\na {{\nabla}}
\def\k {\chi}
\def\bn {\begin{eqnarray}}
\def\en {\end{eqnarray}}
\begin{center}
{\large On exact solutions of a class of fractional Euler-Lagrange
equations}
\end{center}
\vspace{-0.3cm}
\begin{center}
\textbf{Dumitru Baleanu}\footnote{On leave of absence from
Institute of Space Sciences, P.O.BOX, MG-23, R 76900,
Magurele-Bucharest, Romania, Email: baleanu@venus.nipne.ro}\\ Department of Mathematics and Computer Sciences,\\
Faculty of Arts
and Sciences,$\c{C}ankaya$ University- 06530, Ankara, Turkey \\
E-mail: dumitru@cankaya.edu.tr\\ \vspace{0.5cm} \textbf{Juan J.
Trujillo}\\ University of La Laguna, \\Departamento de An\'alisis
Matem\'atico, 38271- La Laguna, Tenerife, Spain\\ Email:
JTrujill@ullmat.es
\end{center}
\begin{abstract}
In this paper, first  a class of fractional differential equations
are obtained by using the fractional variational principles. We
find
 a fractional Lagrangian $L(x(t)$, where $_a^cD_t^\alpha x(t))$
and $0<\alpha< 1$, such that the following is the corresponding
Euler-Lagrange
\begin{equation}
_tD_b^\alpha(_a^cD_t^\alpha) x(t)+ b(t,x(t))(_a^cD_t^\alpha
x(t))+f(t,x(t))=0.
\end{equation}

At last, exact solutions for some Euler-Lagrange equations are
presented. In particular, we consider the following equations
\begin{equation}
_tD_b^\alpha(_a^cD_t^\alpha x(t))=\lambda x(t),\ \ \ (\lambda\in
R)
\end{equation}
\begin{equation}
_tD_b^\alpha(_a^cD_t^\alpha x(t))+g(t)_a^cD_t^\alpha x(t)=f(t),
\end{equation}
where g(t) and f(t) are suitable functions.
\end{abstract}

\vspace{1mm} Keywords: Fractional calculus, differential equations
of fractional order, fractional variational calculus
\newpage

\section{Introduction}
Fractional calculus is an emerging fields and during the last
decades it represents an alternative tool to solve several
problems from various fields \cite{podlubny,zaslavsky,
trujillo,richard,
mainardi1,machado1,machado2,ali,trujillo1,lim,stani}. During the
last years the fractional variational principles
\cite{riewe1,riewe2,klimek1,klimek2,nabulusi,agrawal,agrawalop,agrawal22,
bal,baleanu,baleanu1,baleanu2,baleanu3,bale} have developed and
applied to fractional optimal control problems
\cite{machado,agrawal4}.

Despite of various efforts during the last years, the fractional
Lagrangian and Hamiltonian formulation of both discrete and
continuous systems is at the beginning of its development.
Although the fractional variational principles were started to be
investigated deeply the appropriate physical interpretation of the
fractional derivatives creates  problems in physical
interpretation of the obtained equations. The existence of various
fractional derivatives leads to several Hamiltonian formulations
for a given dynamical system.

Very recently, based on finite difference \cite{jumarie} it was
proposed an alternative definition for the Riemann-Liouville (RL)
derivatives. By using the approach presented in \cite{jumarie} the
troublesome effects of the initial conditions in the RL fractional
derivative are removed.

 By
construction the fractional Lagrangian and fractional Hamiltonian
contain as a particular case the classical counterparts. Due to
the fractional integration by parts, the fractional Euler-Lagrange
equations contains the left and the right Riemann-Liouville
derivatives. Even if the fractional Lagrangian contains only
Caputo derivatives the corresponding fractional Euler-Lagrange
equations contains both RL and Caputo derivatives. From these
reasons we expect to obtain new solutions of the fractional
Euler-Lagrange equations. Another problem which presents interest
is to find a fractional Lagrangian for a given fractional
Euler-Lagrange equations and therefore we obtain a meaning for
these equations. Until now quite a few exact solutions were
reported for fractional Euler-Lagrange equations, therefore
finding more general solutions having physical significance is an
open issue in this area. This issue plays an important role in
fractional quantisation models. Some type of functional involving
the fractional derivatives are used in mathematical economy as
well as utilized for describing the dissipative structures arising
in nonlinear dynamical systems.

The plan of this manuscript is as follows:

 Some basic definitions
of fractional derivatives are shown in section two. Section three
presents the fractional Lagrangian corresponding to a given second
order fractional differential equations involving both RL and
Caputo derivatives. In section four an exact new solution for
fractional oscillator as well as a generalization of it are
obtained.
 Section five is dedicated to our conclusions.
\section{Mathematical tools}
 In this section, we formulate the problem in terms
of the left and the right RL  fractional derivatives, which are
defined as follows,  the~ left~ RL fractional~ derivative
\begin{equation}
{{}_a\textbf{D}_t^{\alpha}f(t)} =
\frac{1}{\Gamma{(n-\alpha)}}\left(\frac{d}{dt}\right)^{n}\int\limits_a^t(t-\tau)^{n-\alpha-1}f(\tau)d\tau,
\end{equation}
and the~ right ~RL ~fractional~ derivative
\begin{equation}
{{}_t\textbf{D}_b^{\alpha}f(t)} =
\frac{1}{\Gamma{(n-\alpha)}}\left(-\frac{d}{dt}\right)^{n}\int\limits_t^b(\tau-t)^{n-\alpha-1}f(\tau)d\tau,
\end{equation}
where the order $\alpha$ fulfills $n-1\leq\alpha <n$ and $\Gamma$
is the Euler's gamma function. If $\alpha$ becomes an integer, we
recovered  the usual definitions, namely,
\begin{equation}
{{}_a\textbf{D}_t^{\alpha}f(t)}
=\left(\frac{df(t)}{dt}\right)^{\a},~~{{}_t\textbf{D}_b^{\alpha}f(t)}
= \left(-\frac{df(t)}{dt}\right)^{\a}; \ \ \ ~(\a=1,2,...).
\end{equation}
Fractional RL derivatives have various interesting properties. For
example the fractional derivative of a constant is not zero,
namely
\begin{equation}
{}_a\textbf{D}_t^{\alpha}C=\ C\
\frac{(t-a)^{-\alpha}}{\Gamma(1-\alpha)}.
\end{equation}
The fractional derivative of a power of t has the following form
\begin{equation}
{}_a\textbf{D}_t^{\alpha}(t-a)^\beta=\
\frac{\Gamma(\alpha+1)(t-a)^{\beta-\alpha}}{\Gamma(\beta-\alpha+1)},
\end{equation}
for $\beta> -1, \alpha\geq 0$. The  Caputo's fractional
derivatives are defined as follows, the left Caputo Fractional
Derivative
\begin{equation}
_a^cD_t^\alpha f(t) = \frac{1}{\Gamma (n-\alpha)} \int_a^t
(t-\tau)^{n-\alpha-1} \left( \frac{d}{d\tau} \right)^n f(\tau)
d\tau ,
\end{equation} 
and the right Caputo Fractional Derivative
\begin{equation}
_t^cD_b^\alpha f(t) = \frac{1}{\Gamma (n-\alpha)} \int_t^b
(\tau-t)^{n-\alpha-1} \left(-\frac{d}{d\tau} \right)^n f(\tau)
d\tau .\end{equation} 
\noindent Here $\alpha$ represents the order of the derivative
such that $n-1 < \alpha < n$.
\section{Fractional variational principles}
Let us consider the following fractional second order differential
equation:
\begin{equation}
{}_t\textbf{D}_{b}^\alpha(_a^cD_t^\alpha) x(t)+ b(t,x(t))(_a^cD_t^\alpha
x(t))+f(t,x(t))=0,
\end{equation}
where $0<\alpha\leq 1$. Our aim is to find a fractional Lagrangian
\begin{equation}
L(x(t),\   _a^cD_t^\alpha x(t)), \hspace{0.2in} 0<\alpha< 1,
\end{equation}
such that
\begin{equation}\label{unu}
\frac{\partial L}{\partial x} +
 _tD_b^\alpha\left(\frac{\partial L}{\partial (_a^cD_t^\alpha x)}\right)={}_t\textbf{D}_{b}^\alpha\left(_a^cD_t^\alpha x\right)+ b(t,x)(_a^cD_t^\alpha
 x)+f(t,x).
\end{equation}

\noindent We assume a solution of this problem as follows
\begin{equation}\label{unu2}
L(x,\  _a^cD_t^\alpha x)=\frac{1}{2}(_a^cD_t^\alpha x)^2 +h(t,x)
_a^cD_t^\alpha x +g(t,x).
\end{equation}
Then, we evaluate the left hand side of (\ref{unu}) and we obtain
\begin{equation}
({}_t\textbf{D}_{b}^\alpha(_a^cD_t^\alpha) x) + (_tD_b^\alpha h(t,x)) +
\frac{\partial{h}}{\partial{x}} (_a^cD_t^\alpha x) +
\frac{\partial{g(t,x)}}{\partial{x}}.
\end{equation}
Therefore, by using (\ref{unu2}) we obtain
 \begin{equation}\label{cv}
\frac{\partial{h(t,x)}}{\partial{x}}=b(t,x),
\end{equation}
and
\begin{equation}\label{cv1}
{}_t\textbf{D}_{b}^\alpha h(t,x) +\frac{\partial{g(t,x)}}{\partial{x}} =f(t,x).
\end{equation}
By using (\ref{cv}) and (\ref{cv1}) we obtain the functions g(t,x)
and h(t,x) respectively.
\section{Exact solutions of fractional Euler-Lagrange equations}
The scope of this section is to present the exact solutions for a
class of problems arising from a fractional variational
principles.
\subsection{One-dimensional fractional oscillator}
The first fractional Euler-Lagrange is given below
\begin{equation}\label{doi}
{}_t\textbf{D}_{b}^\alpha(_a^cD_t^\alpha x(t))=\lambda x(t),
\end{equation}
where $\lambda\in R$.  Our purpose is to solve the equation
(\ref{doi}). For this reason we assume the solution in the
following form
\begin{equation}\label{trei}
x(t)=\sum_{n=0}^{\infty}a_n(t-a)^{n\alpha +\alpha -1},
\end{equation}
where $a_n$ is to be determined. The first step is to calculate
$_a^cD_t^\alpha x(t)$ taking into account (\ref{doi}). Therefore,
we obtain the following
\begin{equation}\label{patru}
_a^cD_t^\alpha x(t)=\sum_{n=1}^{\infty}
a_n\frac{\Gamma((n+1)\alpha)}{\Gamma(n\alpha)}(t-a)^{n\alpha-1}.
\end{equation}
\noindent Then
\begin{equation}\label{cinci}
{}_t\textbf{D}_{b}^\alpha(_a^cD_t^\alpha x(t))=\sum_{n=0}^{\infty}
a_{n+2}\frac{\Gamma((n+3)\alpha)}{\Gamma((n+1)\alpha)}e^{i\pi\alpha}(t-a)^{n\alpha+\alpha-1},
\end{equation}
with $a<x<2b-a$.

Now we find the following relation that permits us to find the
coefficients
 $a_n$, by using (\ref{doi}) and (\ref{cinci})
\begin{equation}\label{sase}
a_{n+2}=\frac{\Gamma((n+1)\alpha)}{\Gamma((n+3)\alpha)}a_n,
\end{equation}
that is the coefficients of the solution of (\ref{doi}) are given
by
\begin{equation}\label{sase}
a_{2(n+1)}=(\lambda
e^{-i\pi\alpha})^{n+1}\frac{\Gamma(\alpha)}{\Gamma((2n+3)\alpha)}a_0,
n \geq 0,
\end{equation}
\begin{equation}\label{sapte}
a_{2(n+1)+1}=(\lambda
e^{-i\pi\alpha})^{n+1}\frac{\Gamma(2\alpha)}{\Gamma(2(n+2)\alpha)}a_1,
n \geq 0.
\end{equation}
So we have
\begin{eqnarray}\label{sapte}
x(t)&=&a_0(t-a)^{\alpha-1}\left[1+\sum_{n=0}^{\infty}\frac{(\lambda
e^{-i\pi\alpha})^{n+1}\Gamma(\alpha)(t-a)^{2(n+1)\alpha}}{\Gamma((2n+3)\alpha)}\right]\cr
&+&a_1(t-a)^{2\alpha-1}\left[1+\sum_{n=0}^{\infty}\frac{(\lambda
e^{-i\pi\alpha})^{n+1}\Gamma(2\alpha)(t-a)^{2(n+1)\alpha}}{\Gamma(2(n+1)\alpha)}\right].
\end{eqnarray}
\par It is obvious to prove that the above series is convergent.
Thus, we obtain the following two general solutions as follows
\begin{eqnarray}\label{opt}
&x_1(t)&=a_0(t-a)^{\alpha-1}[1+\sum_{n=0}^{\infty}(cos(n\pi\
\alpha)\lambda)^{n+1}\frac{\Gamma(\alpha)}{\Gamma(2(n+1)\alpha+\alpha)}(t-a)^{2(n+1)\alpha}]\cr
&+& a_1(t-a)^{2\alpha-1}[1+\sum_{n=0}^{\infty}(cos(n\pi\
\alpha)\lambda)^{n+1}\frac{\Gamma(2\alpha)}{\Gamma(2(n+2)\alpha)}(t-a)^{2(n+1)\alpha}],
\end{eqnarray}
\noindent and
\begin{eqnarray}\label{noua}
&x_2(t)&=a_0(t-a)^{\alpha-1}[\sum_{n=0}^{\infty}(sin(n\pi\
\alpha)\lambda)^{n+1}\frac{\Gamma(\alpha)}{\Gamma(2(n+1)\alpha+\alpha)}(t-a)^{2(n+1)\alpha}]\cr
&+& a_1(t-a)^{2\alpha-1}[\sum_{n=0}^{\infty}(sin(n\pi\
\alpha)\lambda)^{n+1}\frac{\Gamma(2\alpha)}{\Gamma(2(n+2)\alpha)}(t-a)^{2(n+1)\alpha}].
\end{eqnarray}
We observe that for $\alpha=1$, $x_1(t)=a_0cos(t) +a_1 sin(t)$ and
$x_2(t)=0$, therefore the classical result is obtained.
\subsection{A more general case}
In the following we consider the fractional differential equation
\begin{equation}\label{zece}
{}_t\textbf{D}_{b}^\alpha(_a^cD_t^\alpha x(t))+g(t)_a^cD_t^\alpha x(t)=f(t),
\end{equation}
where g(t) and f(t) are suitable functions. \vspace{1mm} We denote $_a^cD_t^\alpha
x(t)=z(t)$ and we rewrite the equation (\ref{zece}) as
\begin{equation}\label{unuu}
{}_t\textbf{D}_{b}^\alpha z(t)+ g(t)z(t)=f(t).
\end{equation}
The equation (\ref{unuu}) can be written as follows
\begin{equation}\label{doii}
\textbf{L}(z(t))=\frac{f(t)}{g(t)},
\end{equation}
where $\textbf{L}=g(t)^{-1}\ _tD_b^\alpha+1$. The solution of (\ref{doii})
can be written in the following form
\begin{equation}
z(t)=\textbf{L}^{-1}\{\frac{f(t)}{g(t)}\},
\end{equation}
where, we will consider
\begin{equation}
\textbf{L}^{-1}=\sum_{i=0}^{\infty}(-1)^{i}[g(t)^{-1}{_tD_b^\alpha}]^{i}.
\end{equation}
The second step is to solve the following equation
\begin{equation}\label{treii}
_a^cD_t^\alpha x(t)=z(t).
\end{equation}
The solution of (\ref{treii}) is as follows
\begin{equation}
x(t)=_aI_t^\alpha z(t) +c_1(t-a)^{\alpha-1} +c_2,
\end{equation}
that is
\begin{equation}
x(t)=\sum_{i=0}^{\infty} (-1)^{i}_aI_t^\alpha
[g(t)^{-1}{_tD_b^\alpha}]^{i}\{\frac{f(t)}{g(t)}\}
+c_1(t-a)^{\alpha-1} +c_2.
\end{equation}
It is very easy to check directly that the above function $x(t)$
is a solution of equation (\ref{zece}), and it is a convergent series if $f(x)$
and $g(x)$ are suitable functions.

\section{Conclusions}
The solutions of the complex fractional Euler-Lagrange equations
were obtained by using the numerical techniques for most of the
cases. In this paper we found a new and more general solution, as
a series solution, of the fractional oscillator within Caputo
derivatives. The classical solution is recovered but a new
solution was also reported. A fractional Lagrangian that produces
 a given class of second order ordinary fractional differential equations was found. By using the operational approach an exact solution
 of a particular Euler-Lagrange equation was obtained.
\section{Acknowledgements}
 The research reported in this paper have been partially supported by
the Scientific and Technical Research Council of Turkey and by MEC
of Spain (MTM2004-0317).


\begin{thebibliography}{99}
\bibitem{podlubny}
Podlubny, I.: \emph{Fractional Differential Equations}, Academic
Press, San Diego CA, (1999).
\bibitem{zaslavsky} Zaslavsky, G. M.: \emph{Hamiltonian Chaos and Fractional Dynamics},
Oxford University Press, Oxford, (2005).
\bibitem{trujillo}
Kilbas,A.A., Srivastava, H. M., Trujillo, J. J.: \emph{Theory and
Applications of Fractional Differential Equations}, Elsevier,
(2006).
\bibitem{richard}
Magin, R. L.: \emph{Fractional Calculus in Bioengineering}, Begell
House Publisher, Inc. Connecticut, (2006).
\bibitem{mainardi1}
Mainardi,F. Luchko, Yu.,Pagnini, G.: The fundamental solution of
the space-time fractional diffusion equation .\emph{Frac. Calc.
Appl. Analys.} \textbf{4(2)} (2001) 153.
\bibitem{machado1} Tenreiro Machado, J.A.:A probabilistic
interpretation of the fractional-order differentiation.Frac. Calc.
Appl. Anal. 8, 73-80 (2003).
\bibitem{machado2}
Tenreiro Machado, J.A.:  Discrete-time fractional-order
controllers. Frac. Calc.  Appl. Anal.4, 47-66 (2001).
\bibitem{ali}
Tofighi, A.: The intrisic damping of the fractional oscillator,
Physica A 329, 29-34 (2006).
\bibitem{trujillo1} Trujillo, J. J.:On a Riemann-Liouville
generalized Taylor's formula. J. Math. Anal. Appl.231, 255-265
(1999).
\bibitem{lim}
Lim , S. C., Muniady, S.V.:Stochastic quantization of nonlocal
fields. Physics Letters A. 324, 396-405 (2004).
\bibitem{stani}
Stanislavsky, A.A.:Fractional oscillator.Phys.Rev.E.,70,051103
(2004).
\bibitem{riewe1} Riewe, F.: Nonconservative Lagrangian and Hamiltonian mechanics,
Phys. Rev. E. 53, 1890-1899 (1996).
\bibitem{riewe2} Riewe, F.: Mechanics with fractional derivatives, Phys. Rev. E 55, 3581-3592 (1997).
\bibitem{klimek1}  Klimek, M.;  Fractional sequential mechanics-models
with symmetric fractional derivatives. Czech. J. Phys.,
51,1348-1354 (2001).
\bibitem{klimek2}
Klimek, M.: Lagrangian and Hamiltonian fractional seqential
mechanics. Czech. J. Phys., 52, 1247-1253 (2002).
\bibitem{nabulusi}
 El-Nabulusi, R.A.: A fractional approach to nonconservative
Lagrangian dynamics. Fizika A, 14(4), 289-298 (2005).

\bibitem{agrawal}Agrawal, O. P.: Formulation of Euler-Lagrange equations for
fractional variational problems.
 J. Math. Anal. Appl. 272, 368-379 (2002).
\bibitem{agrawalop} Agrawal, O.P.:Fractional variational calculus
and the transversality conditions, J. Phys.A. :Math.Gen. 39,
10375-10384 (2006).
 \bibitem{agrawal22}
 Agrawal, O. P.: Generalized Euler-Lagrange equations and
transversality conditions for FVPs in terms of Caputo Derivative,
\emph{in Proc. MME06, Ankara, Turkey, April 27-29, 2006} (Eds. K.
Tas, J.A. Tenreiro Machado and D. Baleanu), to appear in J. Vib.
Control (2007).
  \bibitem{bal}
Rabei, E. M., Nawafleh, K .I., Hijjawi, R. S., Muslih S. I. and
Baleanu D.: The Hamiltonian formalism with fractional derivatives,
J. Math. Anal. Appl.,327, 891-897  (2007).
\bibitem{baleanu}
Muslih, S., Baleanu, D.: Hamiltonian formulation of systems with
linear velocities within Riemann-Liouville fractional derivatives.
J. Math. Anal. Appl. 304(3), 599-603 (2005).

\bibitem{baleanu1}
Baleanu, D.:  Fractional Hamiltoian analysis of irregular systems.
Signal Processing. 86(10),2632-2636 (2006).
\bibitem{baleanu2}
Baleanu, D., Muslih, S.I.: Formulation of Hamiltonian equations
for fractional variational problems. Czech. J. Phys. 55 (6),
633-642 (2005).
\bibitem{baleanu3}
 Baleanu, D., Muslih, S.: Lagrangian formulation of classical
fields within Riemann-Liouville fractional derivatives. Physica
Scripta. 72(2-3), 119-121 (2005).
\bibitem{bale}
 Baleanu, D.,  Avkar, T.: Lagrangians with linear velocities
within Riemann-Liouville fractional derivatives. Nuovo Cimento B.
119, 73-79 (2004).

\bibitem{machado}
Tenreiro-Machado, J.A.:Discrete-time Fractional -order
controllers.Frac. Calc. Appl. Anal., 4(1),47-68 (2001).

\bibitem{agrawal4}
Agrawal, O. P., Baleanu, D.: A Hamiltonian Formulation and a
Direct Numerical Scheme for Fractional Optimal Control Problems,
\emph{in Proc. MME06, Ankara, Turkey, April 27-29, 2006} (Eds. K.
Tas, J.A. Tenreiro Machado and D. Baleanu), to appear in J. Vib.
Control (2007).

\bibitem{jumarie}
Jumarie,G.:Lagrangian mechanics of fractional order,
Hamilton-Jacobi fractional PDE and Taylor's series of
nondifferntiable functions. Chaos, Solitons and Fractals.
32,969-987(2007).


\end{thebibliography}
\end{document}